\documentstyle[aps,epsf,multicol]{revtex}
\begin{document}
\draft
\title{Two-Species Reaction-Diffusion System with Equal Diffusion 
Constants: \\ Anomalous Density Decay at Large Times}
\author{Zoran Konkoli$^{1,2}$ and Henrik Johannesson$^2$}
\address{
  $^1${NORDITA, Blegdamsvej 17, DK 2100 K\o benhavn, Denmark}\\
  $^2${Institute of Theoretical Physics, Chalmers University of Technology 
      and G\"oteborg University, SE 412 96 G\"oteborg, Sweden}}

\maketitle
\begin{abstract}

We study a two-species reaction-diffusion model where $A+A
\longrightarrow \emptyset, A+B \longrightarrow \emptyset$ and $B+B
\longrightarrow \emptyset$, with annihilation rates $\lambda_0,
\delta_0 > \lambda_0$ and $\lambda_0$, respectively. The initial
particle configuration is taken to be randomly mixed with mean
densities $n_{A}(0) > n_{B}(0)$, and with the two species $A$ and $B$
diffusing with the same diffusion constant.  A field-theoretic
renormalization group analysis suggests that, contrary to expectation,
the large-time density of the minority species decays at the same rate
as the majority when $d\le 2$.  Monte Carlo data supports the field
theory prediction in $d=1$, while in $d=2$ the logarithmically slow
convergence to the large-time asymptotics makes a numerical test
difficult.
\end{abstract}
\pacs{82.20.Fd, 05.40.+j}

\begin{multicols}{2}
\narrowtext

Fluctuation effects in low-dimensional reaction-diffusion systems has
received considerable attention lately \cite{review}. When the
dimension $d$ of the system is sufficiently small, below a critical
dimension $d_c$, spatial fluctuations in the particle concentration
dramatically influence the large-time behavior of the system.  Most
strikingly, the large-time scaling exponents and, sometimes,
amplitudes of various observables become universal and independent of
the microscopic description of the system. Indeed, the quest for
possible new universality classes has intensified in recent years,
with studies of a variety of models exhibiting fluctuation-dominated
kinetics at different levels of complexity.  Theoretical advances,
together with a growing number of experimentally accessible
realizations of low-dimensional reaction-diffusion models -- like
reaction kinetics of excitons in polymer chains or surface deposition
of certain proteins \cite{review} -- have further accelerated these
studies.

We here consider a generalization of the well-known single-species
$A+A \rightarrow \emptyset$ and two-species $A+B \rightarrow
\emptyset$ annihilation models \cite{mikhailov} by coupling them
together and allowing {\em all} particles to react:
\begin{equation}
  A+A \stackrel{\lambda_0}{\longrightarrow}\emptyset   \ \ \ \
  A+B \stackrel{\delta_0 }{\longrightarrow}\emptyset   \ \ \ \
  B+B \stackrel{\lambda_0'}{\longrightarrow}\emptyset . 
  \label{eq1}
\end{equation}
The $A$ and $B$ particles are assumed to perform a random walk on a
d-dimensional lattice with diffusion constants $D_A$ and $D_B$
respectively. When two particles meet on the same site they annihilate
with probabilities given by the reaction rates $\lambda_0$,
$\delta_0$, and $\lambda_0'$ in (\ref{eq1}). Among a number of
possible applications, the model may be used to describe {\em steric}
reaction-diffusion processes \cite{JChemPhys}, with the two species
representing the projections of the chemically ``active spots'' of the
reactants onto some fixed reference direction \cite{we}.

The question we wish to address is how the average particle densities
$n_A(t)$ and $n_B(t)$ decay in time when there is an initial imbalance
$n_{A}(0) > n_{B}(0)$ in the population of the two species. (In an
application to a steric reaction-diffusion process, this would
correspond to a bias in the initial orientations of the reactants,
triggered by some external field \cite{JChemPhys}.) When $\delta_0 >
\lambda_0, \lambda_0'$ we expect the minority species of $B$-particles
to die out before the majority $A$. This has been confirmed by a
field-theoretic renormalization group analysis for the case when
$D_A\ne D_B$ \cite{howard}: Independent of dimensionality, the
minority species was shown to vanish before the extinction of the
majority. Here we shall focus on the case when $\lambda_0' =
\lambda_0$ and with the diffusion constants tuned to the same value,
$D_A = D_B \equiv D$, as would be the proper choice when applying the
model to a steric reaction-diffusion process \cite{JChemPhys}. Since
the {\em effective} time-dependent reaction rates are controlled by
the diffusion constants \cite{mikhailov}, it is interesting to explore
whether the case of {\em equal} diffusion constants could lead to new
effects.

To set the stage, let us briefly study the classical rate equations
implied by (1):
\begin{equation}
  \frac{da}{dt} = -(\lambda_0 a^2 + \delta_0 ab),
  \ \ \ \ 
  \frac{db}{dt} = -(\lambda_0 b^2 + \delta_0 ba),
  \label{eq2}
\end{equation}
with $a, b$ the mean-field densities of species $A, B$, and with
initial conditions $a(0)=n_{A}(0)$ and $b(0)=n_{B}(0)$.

The qualitative behavior of (\ref{eq2}) can be studied by a mapping to
a Poincar\'{e} sphere \cite{ODE} with $z = 1/a$ and $u=b/a$:
\begin{equation}
  \frac{dz}{dt}=\lambda_0 + \delta_0 u(t), \ \ \ 
  \frac{du}{dt}=(\lambda_0 - \delta_0) u (1-u) / z \ ,
  \label{eq3}
\end{equation}
where u is assumed to be in the interval $[0,1]$.  With $\lambda_0 <
\delta_0$, (\ref{eq3}) implies that for $u(0)<1$ and for large times
$1/u(t)$ diverges as
\begin{equation}
  \frac{1}{u(t)}= 
    \frac{a(t)}{b(t)}\propto t^{\delta_0/\lambda_0-1} 
    \rightarrow \infty
  , \ \ \ 
  t \rightarrow \infty,
  \label{eq4}
\end{equation}
when $a(0) > b(0)$. Thus, with $\lambda_0$ and $\delta_0$ fixed, the
minority species is killed off early, as expected. Note, however, that
(\ref{eq4}) suggests that $u(t)$ could saturate to a constant if the
effective {\em diffusion-controlled} reaction rates replacing
$\lambda_0$ and $\delta_0$ in the presence of fluctuations approach
each other asymptotically.  To explore this possibility it is
convenient to pass to a field-theoretic formulation of the problem.

For this purpose, let us consider the master equation of the system: 
\begin{equation}
  \frac{d}{dt} P(c,t) = \sum_{c'}R_{c'\rightarrow c} P(c',t)  
  - \sum_{c'}R_{c \rightarrow c'} P(c,t). 
  \label{eq5}
\end{equation}
Here $P(c,t)$ denotes the probability of a configuration $c$ at time
$t$, and $R_{c' \rightarrow c}$ is the transition rate from state $c'$
into $c$, determined by $D, \delta_0$ and $\lambda_0$. We take
$P(c,0)$ as a Poisson distribution, with averages denoted by
$n_{A}(0)$ and $n_{B}(0)$ for $A$ and $B$ particles
respectively. Eq.~(\ref{eq5}) can be translated into a
Schr\"odinger-type equation with a second-quantized Hamiltonian
\cite{DoiPeliti}, which in the continuum limit turns into a field
theory with action $S$:
\begin{eqnarray}
 && S = \int d^dx \int dt 
      [ \sum_{\alpha=A,B}
           \bar{\phi}_\alpha(\partial_t-\nabla^2)\phi_\alpha +
       \sum_{\alpha,\beta=A,B}
         \lambda_{\alpha\beta}\times
    \nonumber \\
 && \times  \bar{\phi}_\alpha (1+\bar{\phi}_\beta/2)
       \phi_\alpha \phi_\beta 
    ] 
    - \int d^dx \sum_{\alpha=A,B} n_\alpha(0) \bar{\phi}_\alpha(x,0) ,  
    \label{Spp}
\end{eqnarray}
with rescaled time variable, $Dt \rightarrow t$.  The scalar fields
$\phi_A$ and $\phi_B$ describe the A and B particles with initial
densities $n_{\alpha}(0)$, $\alpha=A,B$, and with the coupling
constants $\lambda_{AA}=\lambda_{BB}=\lambda_0$ and
$\lambda_{AB}=\lambda_{BA}=\delta_0$ encoding the reaction rates.  The
action in (\ref{Spp}) faithfully mimics the master equation in
(\ref{eq5}), up to terms irrelevant under renormalization.  Note that
the rate constants become dimensionless when $d=2$ which suggests that
$d_c=2$, a result confirmed by standard power counting
\cite{amit}. Since we are interested in effects from fluctuations, we
focus on $d \le 2$ in what follows.

When $d \le 2$ the particle densities can be calculated perturbatively
from (\ref{Spp}) as $n_{\alpha}(t) = <\!\phi_{\alpha}\!>_S$ using an
$\epsilon-$expansion, with the rate constants $\lambda_0$ and
$\delta_0$ replaced by renormalized rates $g_{\alpha}(\kappa^{-2})
=\kappa^{-\epsilon} \alpha_0/(1+\kappa^{-\epsilon}\alpha_0/g^*)$, with
$\alpha = \lambda$ and $\delta$, respectively, and with $\kappa$ the
renormalization scale. Divergent logarithmic terms in the expansion
are conveniently grouped together using ``running coupling constants''
\begin{equation} 
  g_{\lambda}(t) = 
   \frac{g^*}{1+(t/t_\delta^*)^{-\epsilon/2}}, \ \ \ g_{\delta}(t) =
   \frac{g^*}{1+(t/t_\lambda^*)^{-\epsilon/2}}
   \label{run}
\end{equation}
where $g^* = \Gamma(\epsilon/2)^{-1}(8\pi)^{d/2}$ is the
renormalization group (RG) fixed point to which $g_{\lambda}(t)$ and
$g_{\delta}(t)$ flow under a change of time scale \cite{warning}.  The
running coupling constants in (\ref{run}) are obtained from the
Callan-Symanzik equation which expresses the independence of the
densities $n_\alpha(t)$ on the choice of $\kappa$:
\begin{eqnarray}
  && n_\alpha(t;g_{\lambda}(\kappa^{-2}),g_{\delta}(\kappa^{-2});
        n_{A,0},n_{B,0};\kappa)=
     (\kappa^2t)^{-d/2} \times \nonumber \\ 
  && \times n_\alpha(\kappa^{-2};g_{\lambda}(t),g_{\delta}(t);
          \tilde{n}_{A,0}(t),\tilde{n}_{B,0}(t);\kappa)  \:,\: \
    \alpha=A,B  
  \label{CSAB}
\end{eqnarray}
with $\tilde{n}_{\alpha,0}(t)=(\kappa^2t)^{d/2}n_{\alpha,0}$.

Employing the mean-field densities from (\ref{eq2}) in (\ref{CSAB})
{\em (fluctuation-improved mean-field treatment} \cite{howard,lee}),
we obtain
\begin{equation}
  u(t) \rightarrow u^* = (\Omega-1)/(\Omega+1)
  \label{u} \ ,
\end{equation}
where $\Omega^2=1+4\xi^* u_0/(1-u_0)^2$ and $\xi^*=
\mbox{exp}(-8\pi(\frac{1}{\lambda_0} - \frac{1}{\delta_0}))$, $1$ for
$d=2$, $1$.  Thus, at this level of analysis the minority species
at large times decays at the same rate as the majority ($u^* \neq 0$). Note
also that (\ref{u}) implies that $u(d=1,t)>u(d=2,t)$ as
$t\rightarrow\infty$, suggesting that the survival rate of the
minority particles increases with reduced dimensionality.

The fluctuation-improved mean-field analysis (where all tree-level
diagrams are summed) takes into account effects from fluctuations via
the renormalized coupling constants. However, in order to fully
account for fluctuations one has to explicitly consider the loop
diagrams.  For this purpose it is convenient to expand in the
asymptotically small parameter $\eta(t)\equiv
g_{\lambda}(t)-g_{\delta}(t)$. To one-loop level and to first order in
$\eta$ we thus obtain
\begin{equation}
  n_\alpha(t) = \mu_\alpha N(t)(1+\Delta_\alpha(t)) \ , 
   \ \ \   \alpha = A,B \label{ad1}. \\
  \label{abd}
\end{equation}
Here $\mu_A=1/(1+u^*)$, $\mu_B=u^*/(1+u^*)$ and $\Delta_A(t),
\Delta_B(t) \propto \eta(t) \ln t^{d/2} \rightarrow 0$ as $t
\rightarrow \infty $, and we hence recover the result in (\ref{u}) at
very large times. The common factor $N(t)$ in (\ref{abd}) coincides
with the density of the single species $A+A {\longrightarrow}\
\emptyset $ model, with $N(t)=\ln t/8\pi t$ in $d=2$ and $N(t)=
1/\sqrt{8\pi} t^{-1/2}$ in $d=1$ \cite{lee}. The approach of the total
density $n_A(t) + n_B(t)$ to that of the single-species model reflects
the fact that the equal values of the diffusion-controlled rates in
(\ref{run}) at large times makes the two species indistinguishable,
effectively leaving us with a single-species system. We should here
stress that the values of the amplitudes in (\ref{abd}) have been
obtained by ignoring all irrelevant terms (in RG sense) that could be
added to the action in (\ref{Spp}). Since we do not expect the
amplitudes to be universal, such terms - if included - could shift the
value of $u^*$ in (\ref{u}). The scaling form of $n_A(t)$ and $n_B(t)$
on the other hand {\em is} expected to be universal (considering the
generic behavior of this class of problems \cite{mikhailov}), and
should be insensitive to any left-out irrelevant terms.

To summarize: Our RG analysis to one-loop level suggests that the
minority species at large times decays at the same rate as the
majority when $d \le2$.  This result - if robust against contributions
from higher-order terms in the loop expansion - is quite remarkable
and very different from the conventional picture implied by classical
rate theory. Unfortunately, due to the difficulty of keeping track of
possible ``correction-to-scaling'' terms \cite{amit}, the present model
(in contrast to the simpler $A+A \rightarrow \emptyset$ model) is
intractable to standard procedures for assessing higher-loop effects
\cite{unpub}.  Therefore, to check the robustness of the result in
(\ref{abd}) we have resorted to a Monte Carlo simulation of the master
equation (\ref{eq5}) (which in addition provides information on the
transient kinetics).

To perform the simulation we have used an {\em improved minimal
process algorithm} \cite{numerics}: At each time step a lattice site,
say $i$, is picked at random (out of L possible sites), and a table of
statistical weights $w^{(i)}_\alpha$ for the possible processes
$\pi_\alpha$, $\alpha=1, \ldots, M$, at this site is constructed, with
$\pi_M$ a ``null process''.  The weights are defined as cumulative
probabilities: $w^{(i)}_\alpha = \sum_{\beta=1}^{\alpha} R^{(i)}_\beta
/ Q $, with $Q=\sum_{\alpha=1}^M R^{(i)}_\alpha$, where
$R^{(i)}_\alpha$ denotes the rate for $\pi_\alpha$. $Q$ is the total
rate which is kept the same for all sites by properly adjusting the
rates $R^{(j)}_M$ $j=1, \ldots, L$ for the null processes.  After
selecting a random number $0\le r<1$ that process $\pi_\alpha$ is
carried out for which $w^{(i)}_{\alpha-1}\le r<w^{(i)}_\alpha$
($w^{(i)}_{-1}\equiv 0$), and time is incremented by $t\rightarrow
t+1/LQ$.

\begin{figure}
\epsfxsize=9cm
\epsfysize=8cm
\epsfbox{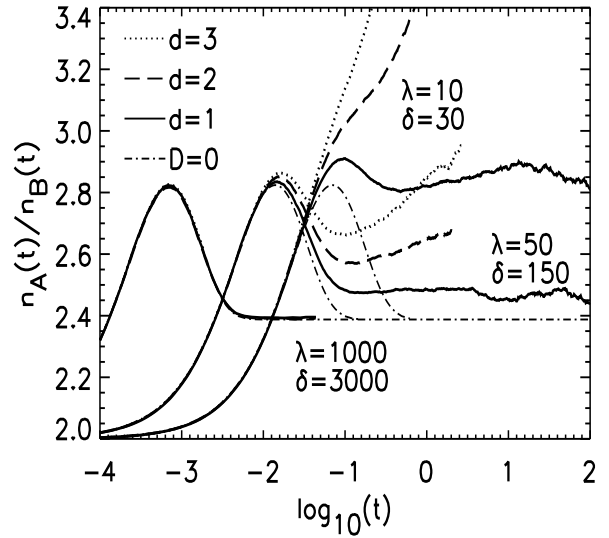}
\caption{Density ratio $n_A(t)/n_B(t)$ obtained from a numerical
simulation of the master equation (\ref{eq5}) in $d=1$, $2$ and $3$ on
a $10^6$, $10^3\times 10^3$ and $10^2\times 10^2\times 10^2$ lattice,
and numerical solution of (\ref{eq5}) in the limit of $D=0$
(``decoupled-sites problem'').  The fixed parameter values are:
$D=1s^{-1}$, $n_{A,B}(0)=2,1$ particles/site. The reaction rates
$\lambda_0$ and $\delta_0$ are given in units of the diffusion
constant $D$.}
\label{fig1}
\end{figure}

Fig. 1 shows results for the density ratios $n_A(t)/n_B(t)$ from
simulations in 1, 2 and 3 dimensions. Initially the system is
homogeneous with randomly mixed A- and B-particles, with $n_A(0) >
n_B(0)$.  At early times a typical minority particle is then
surrounded by several A's, and is therefore likely to be killed
off. As seen in Fig.~\ref{fig1}, this is indeed the case: in all
dimensions the density ratio grows initially.  However, for
sufficiently large rates there is a {\em recovery effect}: After the
initial growth the density ratio drops --- the minority species finds
a way to recover.  (Here and in the following ``recovery'' is
understood in relative terms. The number of minority particles still
decays in time but the ratio of majority to minority densities
decreases in favor of the minority). As seen in Fig. 1 the large-time
behavior of the curves depend on the dimensionality as well as on the
values of the reaction rates. For $\delta = 3\lambda_0 =50$ the $d=3$
curve takes off after the recovery phase, in agreement with the
classical mean field result (\ref{eq4}). This is in striking contrast
to the $d=1$ curve which saturates to a plateau, as predicted by the
one-loop RG result in (\ref{abd}). The eventual fate of the $d=2$
curve is more difficult to foretell since the convergence to the
asymptotic scaling regime in $d=2$ is logarithmically slow
\cite{unpub}. Unfortunately, with available techniques it is hard to
push the $d \ge 2$ simulations reported in Fig. 1 much beyond
log$_{10}(t) \approx 0.5$ where the number of particles get too small
to obtain a reliable statistics.  Turning to the $\delta = 3\lambda_0
=10$ runs, the $d=1$ curve again saturates to a plateau, while the
$d=2$ and $d=3$ curves take off even before a peak has developed. For
this case it may be harder to argue that the $d=2$ curve may
eventually also saturate, although it cannot be excluded considering
the slow convergence.  Note that the RG prediction of an enhanced
survival rate in low dimensions is confirmed for this choice of
reaction rates as well: the $d=1$ curve falls below that for
$d=2$. The runs with very large reaction rates, $\delta = 3\lambda_0
=10^3$, produce curves which collapse onto a single curve, spuriously
suggesting a plateau in all dimensions: In $d=3$, however, large
reaction rates impede the crossover to the mean field asymptotics, and
we expect the $d=3$ curve to take off at a time not accessible with
our simulation. On the other hand, we do expect the $d=1$ curve to
stay flat (considering the generic behavior for smaller rates), while
the asymptotics of the $d=2$ curve is again hard to predict due to the
logarithmically slow convergence to the scaling region.

\begin{figure}
\epsfxsize=9cm
\epsfysize=8cm
\epsfbox{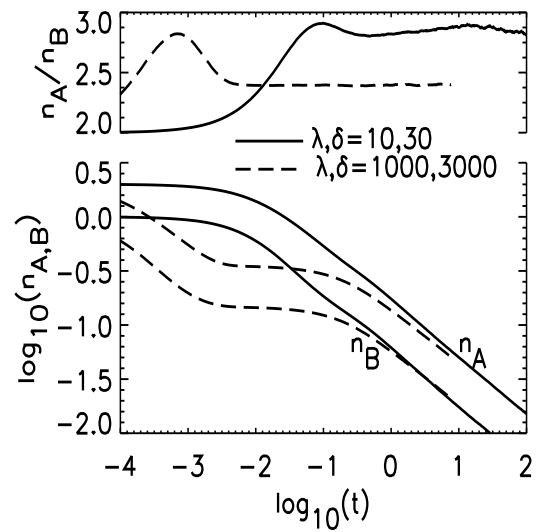}
\caption{Plot of individual densities (bottom) together with density
ratios $n_A(t)/n_B(t)$ (top) in $d=1$ obtained from simulations.}
\label{fig2}
\end{figure}

The build-up of a density-ratio peak with a subsequent ``recovery
phase'' in the early stage of the kinetics (for sufficiently large
rates) is made transparent by solving the master equation (\ref{eq5})
with the diffusion constant $D$ set to zero (i.e. with an ensemble of
decoupled single site problems \cite{footnote}).  The hatched curves
in Fig.~\ref{fig1} represent the corresponding density ratios for very
large and small reaction rates.  Initially the dominant $A+B
\longrightarrow \emptyset$ reactions on the different sites cause a
fast rise of $n_A(t)/n_B(t)$ {\em (peak formation)}. After some time
this process gets exhausted with single B-particles left at a finite
fraction of the sites. The $A+A \longrightarrow \emptyset$ reactions
on the remaining sites now become dominant, causing a depletion of
A-particles {\em (recovery phase} for B particles). As this process in
turn gets exhausted one is left with a constant ratio $n_A(t)/n_B(t)$
{\em (plateau}).  As seen in Fig.~\ref{fig1}, for small rates the
early-time decoupled-sites kinetics gets perturbed by diffusion of
particles from neighboring sites, which leads to an enhanced mixing of
A and B particles and hence a steeper rise of the density-ratio
curve. Still, in $d=1$, the curve eventually saturates to a
plateau. This plateau, however, is of a very different origin from
that of the decoupled-sites problem with no diffusion. This is
strikingly shown in Fig.~\ref{fig2} where we have plotted the
individual densities in $d=1$ for large and small rates, with the
corresponding density ratio curves inserted at the top.  In both cases
the individual densities are seen to decay {\em after the plateau has
formed}. Note in particular that the decoupled-sites plateau for large
rates (with no decay of the individual densities) persists only in the
time interval $-2.5 \le $log$_{10}(t) \le -1$. At later times effects
from diffusion sets in, without, however, visibly distorting the
plateau. Note also that although the asymptotic curves come very close
to each other, they do not collapse onto a single curve, in agreement
with the one-loop RG result: The total density $n_A(t) + n_B(t)$ in
(\ref{abd}) is universal, whereas the individual components are not.

\begin{figure}
\epsfxsize=9cm
\epsfysize=8cm
\epsfbox{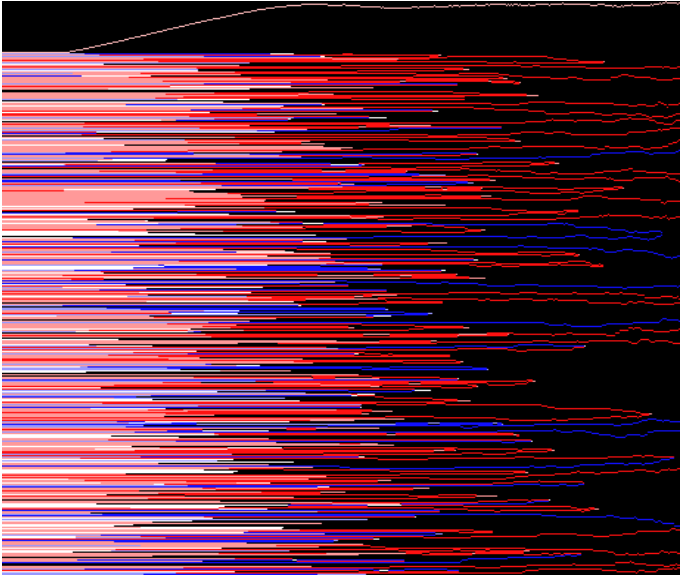}
\caption{``Cartoon'' of the $d=1$ kinetics. Time runs horizontally,
with 300 lattice sites represented along the vertical axis.  At the
top the corresponding density ratio is plotted for comparison. The
reaction rates are $\lambda_0,\delta_0=10,30$. Red (blue) color
denotes that a majority of particles belong to the A (B)
species. White indicates the same number of A's and B's on a lattice
site, with intermediate cases covered by intermediate colors. Black
color denotes an empty site.}
\label{fig3}
\end{figure}

It is instructive to look at a Monte Carlo ``cartoon'' of the kinetics
in $d=1$, as plotted in Fig.~\ref{fig3}. The ``recovery phase'' (for B
particles) is visible as dense red ropes of A particles terminating in
black (empty sites), with the remaining particles diffusing as single
threads colored red (A particles) or blue (B particles). Remarkably,
at large times these threads self-organize spatially in such a way as
to keep the density ratio at a constant value. By a comparison to the
$A+B \rightarrow 0$ model \cite{unpub} one observes that the tendency
for domain formation in the present model is much weaker: A and B
particles appear to be rather well mixed asymptotically. This makes
the anomalous density decay even more intriguing, suggesting that the
average A-A and A-B distances are correlated asymptotically so as to
precisely compensate for the smaller number of A-B pairs (with
endangered B particles) as compared to A-A pairs.

To conclude, we have studied a two-species reaction-diffusion model
where the particle densities exhibit an anomalous behavior at large
times: The spatial self-ordering of the two species allows the
minority population to persevere, with the asymptotic density ratio
(of majority to minority species) kept at a constant value. This
apparent ``non-Darwinian'' kinetics is entirely driven by fluctuations
in the diffusion-controlled regime, typical for a system of reduced
dimensionality.  A perturbative RG analysis suggests that the same
remarkable behavior is present also in two dimensions, although this
is still to be verified by independent methods. Details and extensions
will be published elsewhere \cite{unpub}.

It is a pleasure to thank B.~P.~Lee for stimulating discussions and
for generously sharing of his insights.  We also wish to thank
K.~Oerding, P.~A.~Rey, M.~Howard, G.~M.~Sch\"utz, M.~Henkel, and
I. Sokolov for helpful discussions. H.J. acknowledges support from the
Swedish Natural Science Research Council.

\end{multicols}


\begin{thebibliography}{19}
\bibitem{warning} We should here warn the reader that the time
dependent rates in (7) carry a different meaning from those in the
more conventional Smoluchowski approach \cite{mikhailov}.
\bibitem{footnote} {The decoupling can also be achieved by sending the
reaction rates to infinity, keeping $D$ fixed. The two procedures are
equivalent, up to a rescaling of the time variable.}
\end{thebibliography}
\end{document}